\documentstyle[preprint,pre,aps,psfig]{revtex}
\tighten
\def\be{\begin{equation}} 
\def\ee{\end{equation}}
\def\pa{\partial}
\def\na{\nabla}
\def\lan{\langle}
\def\ran{\rangle}
\def\pr{\prime}
\def\rarrow{\rightarrow}
\def\iff{\infty}

\begin{document}

\title{In search of smooth sandpiles:\\ the Edwards-Wilkinson equation
with flow}

\author{Parthapratim Biswas\footnote{ppb@boson.bose.res.in},
Arnab Majumdar\footnote{arnab@boson.bose.res.in}, 
Anita Mehta\footnote{anita@boson.bose.res.in}}

\address{ S.N.Bose National Centre For Basic Sciences\\
Salt Lake City, Block JD, Sector III, Calcutta-700 091, INDIA}

\author{J.K.Bhattacharjee\footnote{tpjkb@iacs.ernet.in}}

\address{ Indian Association for the Cultivation of Science\\
Jadavpur,Calcutta-700 032, INDIA}

\maketitle

\begin{abstract}
The well-known Edwards-Wilkinson equation with a flow term
added  exhibits a smoothing fixed point in addition
to the normal EW fixed point. Based on this, we present
a model of sandpiles involving a coupling between fixed and mobile 
grains, which then shows the smoothing behaviour
that would be expected to obtain after avalanche propagation.
 
\end{abstract}
\draft

The
dynamics of sandpiles have 
intrigued researchers in physics
over recent years \cite{kn:rpp,kn:ambook}
with a great deal of effort being devoted to
the development of techniques involving for instance
cellular automata \cite{kn:btw,kn:ca}, continuum
equations \cite{kn:ambook,kn:mln,kn:amrjnsd} and Monte Carlo schemes
\cite{kn:prl} to investigate this very complex
subject. However what have often been lost sight of
in all this complexity are some of the extremely
simple phenomena that are exhibited by granular media
that still remain unexplained.

One such phenomenon is that of the smoothing of
a sandpile surface after the propagation of an avalanche
\cite{kn:sdnagel}. It is clear what happens physically:
an avalanche provides a means of shaving off roughness
from the surface of a sandpile by transferring grains
from bumps to available voids \cite{kn:ca}, and thus
leaves in its wake a smoother surface. However, surprisingly,
researchers have not to our knowledge come up with
a model of sandpiles that has exhibited this behaviour.

In this paper we present therefore a very simple
model involving coupled continuum equations
which exhibits smoothing. In work that is presently
under completion \cite{kn:garywidth},
we have constructed a cellular automaton model
of a sandpile surface where this smoothing behaviour
is also made manifest.

This paper comprises : (i) an introductory section,
where the equations are presented and explained
(ii) a section where numerical and analytical results
are obtained 
(iii) a concluding section where the physics of our results is discussed.

\section {Introduction: the equations}

Coupled equations  have been seen to be immensely helpful
in representing the dynamics of systems where there are
{\it two} clear relaxational mechanisms, for instance
 those corresponding respectively to particles 
moving independently of each other  and  
to their collective motion within clusters \cite{kn:ambook,kn:amrjnsd}.
In the case of sandpiles, 
these mechanisms correspond respectively 
to the intercluster and intracluster motion of individual grains \cite{kn:mln}.

The model we present here involves just such
a pair of coupled equations, where the equation
governing the evolution of clusters (``stuck" grains)
is closely related to the very well-known Edwards-Wilkinson (EW)
model \cite{kn:ew}. The equations are:
\begin{eqnarray}
{\pa h\over \pa t}             & =& D_h\na^{2}h +c\na h+\eta(x,t)  \nonumber\\
{\pa\rho\over\pa t}            & =& D_{\rho}\na^{2}\rho -c\na h   
\label{eq-ew}
\end{eqnarray}
where the first of the equations describes the height $h(x,t)$
of the sandpile surface at $(x,t)$ measured from some mean $\langle h \rangle$,
and is precisely the EW equation in the presence of the flow
term $c\na h$. The second equation describes the evolution of
flowing grains, where $\rho(x,t)$ is the local density of such grains
at any point $(x,t)$. As usual, the noise $\eta(x,t)$ is taken
to be Gaussian so that:
$$ \lan \eta(x,t)\eta(x^{\pr},t^{\pr})\ran=\Delta^{2}\delta(x-x^{\pr})\delta(t-t^{\pr}).$$
with $\Delta $ the strength of the noise. Here, $\langle \cdots \rangle$ refers
to an average over space as well as over noise.

Before further analysis, we review some general facts about rough interfaces
\cite{kn:interf}.
Three critical exponents, $\alpha$, $\beta$, and $z$,
characterise the spatial and temporal scaling
behaviour of a rough interface.
They are conveniently defined by considering the (connected)
two-point correlation function of the heights
\be
S({\bf x}-{\bf x}^{\prime},t-t^{\prime})=\bigl\langle h({\bf x},t)h({\bf x}^{\prime},t^{\prime})\bigr\rangle
-\bigl\langle h({\bf x},t)\bigr\rangle\bigl\langle h({\bf x}^{\prime},t^{\prime})\bigr\rangle.
\ee
We have
\be
S({\bf x},0)\sim \vert{\bf x}\vert^{2\alpha}\quad(\vert{\bf x}\vert\to\infty)
\quad  \mbox{and} \quad S({\bf 0},t)\sim \vert t\vert^{2\beta}\quad(\vert t\vert\to\infty),
\ee
and more generally
\be
S({\bf x},t)\approx\vert{\bf x}\vert^{2\alpha}F
\bigl(\vert t\vert/\vert{\bf x}\vert^z\bigr)
\ee
in the whole long-distance scaling regime (${\bf x}$ and ${\bf t}$ large).
The scaling function $F$ is universal in the usual sense;
$\alpha$ and $z=\alpha/\beta$ are respectively referred to
as the roughness exponent and the dynamical exponent of the problem.

In addition, we have for the full structure factor which is the 
double Fourier transform $S(k,\omega)$
\be
S(k,\omega) \sim \omega^{-1}k^{-1-2\alpha}\Phi(\omega/k^{z} )
\ee
which gives
in the limit of small $k$ and $\omega$,  
\be
S(k,\omega = 0) \sim k^{-1-2\alpha -z} \hskip 0.4cm ( k\rarrow 0)  \quad \mbox{and} \quad
S(k=0,\omega) \sim \omega^{-1-2\beta - 1/z} \hskip 0.4cm ( \omega\rarrow 0)
\label{eq-kw1}
\ee
The scaling relations for the corresponding single Fourier transforms are
\be
S(k,t = 0) \sim k^{-1-2\alpha} \hskip 0.4cm ( k\rarrow 0) \quad \mbox{and} \quad
S(x=0,\omega) \sim \omega^{-1-2\beta} \hskip 0.4cm ( \omega\rarrow 0) \label{eq-1d}
\ee

\section {Numerical and analytical results}
\subsection{Analysis of the decoupled equation: the Edwards-Wilkinson equation with flow}
For the purposes of analysis, we focus on the first of the
two coupled equations  (Eq.\ref{eq-ew}) presented above,
\be
{\pa h\over \pa t}= D_h\na^{2}h +c\na h+\eta(x,t)
\label{eq-few}
\ee
 noting that this equation 
is essentially decoupled from the second. (This statement
is, however, not true in reverse, which has implications
to be discussed later). 
We note that this is entirely equivalent to the
 Edwards-Wilkinson equation \cite{kn:ew} in a frame moving with velocity $c$
$$ x^{\pr} = x+ct \, , \quad  t^{\pr} = t $$
and would on these grounds expect to find only the
well-known EW exponents $\alpha=0.5$ and $\beta=0.25$ \cite{kn:ew}.
This would be verified by naive single Fourier transform analysis 
of Eq.\ref{eq-few} which yields these exponents via 
Eq.\ref{eq-1d}.

 Eq.\ref{eq-few} 
can be solved exactly as follows. 
Using the standard form
$$ {\it \hat{L}} h = \eta(x,t) $$
with 
$$ {\it \hat{L}} = \big [ {\pa \over \pa t }-D_h\na^2 - c\na \big ] $$ 
we find for the propagator $G(k,\omega)$
$$ G = (-i\omega+D_h k^2+i kc )^{-1} $$
This can be used to evaluate the structure factor
\be  
S(k,\omega) = {\langle h(k,\omega)h(k^{\pr},\omega^{\pr}) \rangle \over 
\delta(k+k^{\pr})\delta(\omega+\omega^{\pr}) }
\ee
which is the Fourier transform of the full correlation function:

\be
 G(x-x^{\pr},t-t^{\pr}) = {\lan h(x,t)h(x^{\pr},t^{\pr}) \ran}.
\ee
The solution for $S(k,\omega)$ so obtained is: 
\be
S(k,\omega)  =  {\Delta^2 \over (\omega-ck)^{2}+D_h^{2}k^{4}} 
\ee

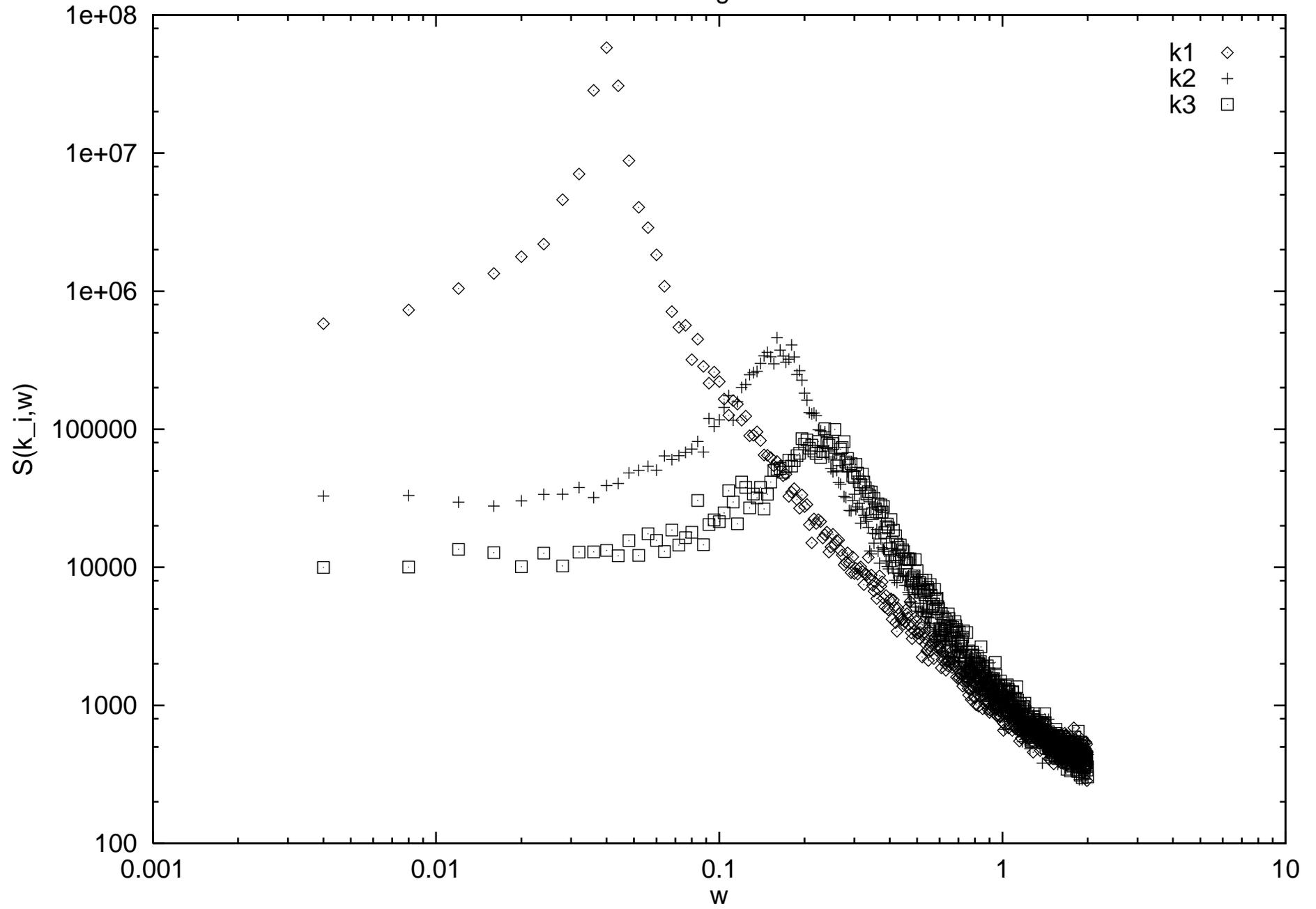
\begin{figure}
\caption{
  The double Fourier transform,
$S(k=k_i,\omega)$
for the $h$-$h$ correlation function for three wavevectors
$k_1 = 0.02$, $k_2 = 0.08$ and 
$k_3 = 0.12$. ($c = 2$ and $D_h = \Delta^2 = 1.0$) 
}
\label{fig-1}
\end{figure}

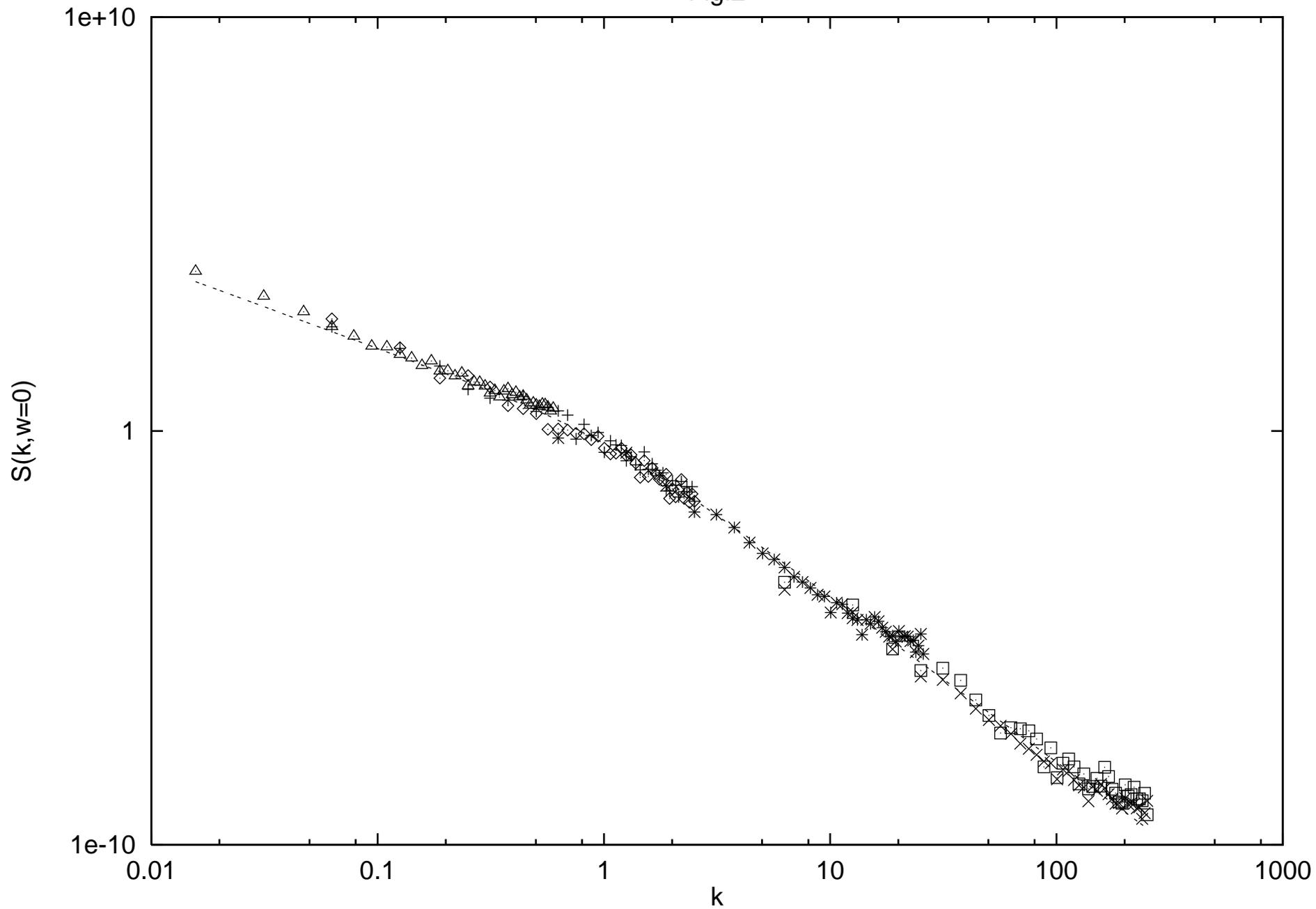
\begin{figure}
\caption{
The double Fourier transform,
$S(k,\omega=0)$ 
for the $h$-$h$ correlation function, along
with a  fit to 
$\left({k^{2.00\pm0.03}+k^{4.00\pm0.05}}\right)^{-1}$.
($c = D_h = \Delta^2 = 1.0$)
}
\label{fig-2}
\end{figure}

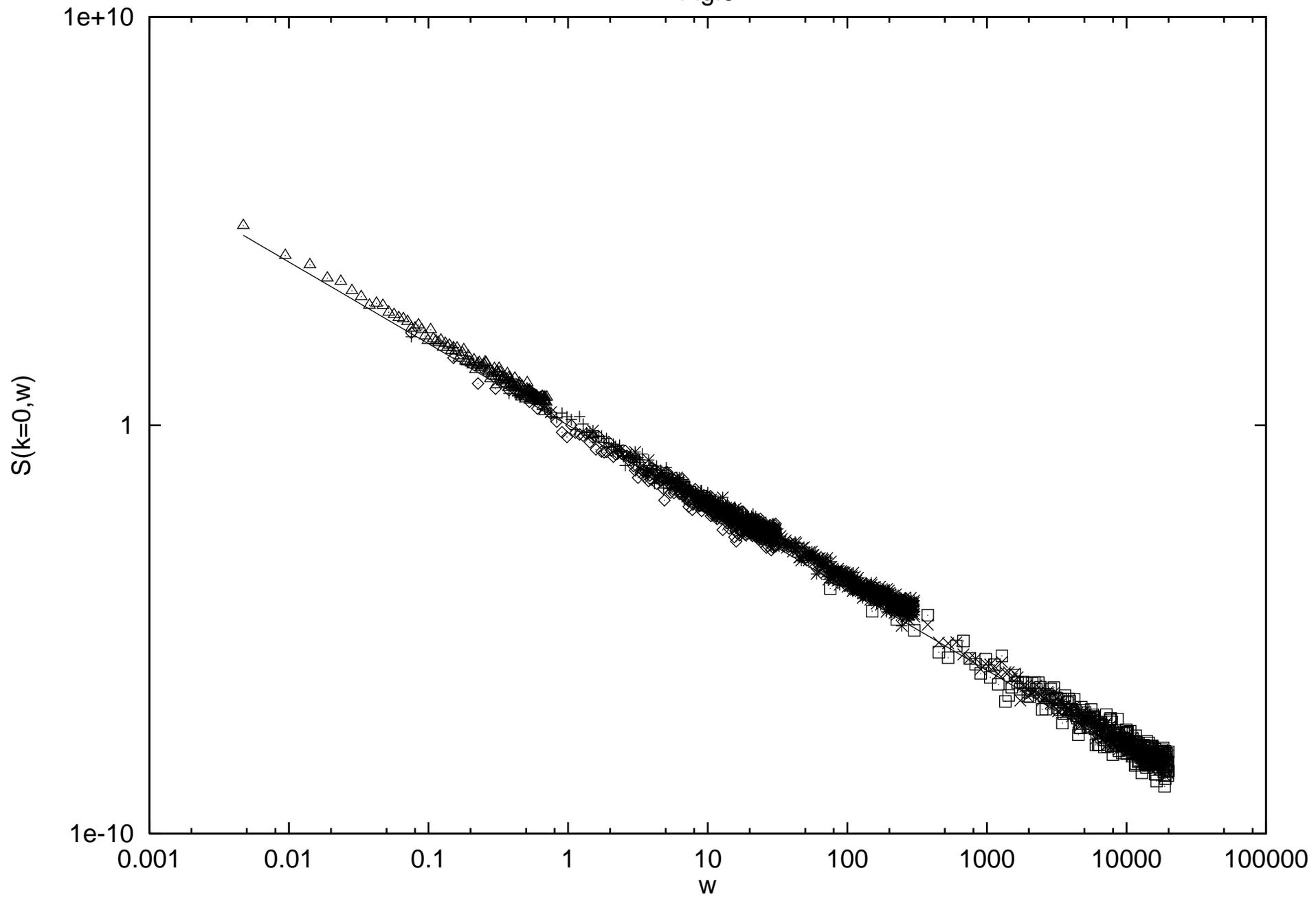
\begin{figure}
\caption{ 
The double Fourier transform,
$S(k=0,\omega)$ 
for the $h$-$h$ correlation function, along
with a  fit to 
$\omega^{-2.00\pm 0.05 } $.
($c = D_h = \Delta^2 = 1.0$)
}
\label{fig-3}
\end{figure}

This is illustrated in Fig.\ref{fig-1} while 
representative graphs for 
 $ S(k,\omega=0)$ and 
 $S(k=0,\omega)$ are presented in Figs.\ref{fig-2} and \ref{fig-3} respectively.
It is obvious from the above that $S(k,\omega)$ does not show simple scaling. 
More explicitly, if we write 
$$ S^{-1}(k,\omega=0) = \frac{\omega_0^2}{\Delta^2} \left( \frac{k}{k_0} \right)^2
\left[ 1 + \left( \frac{k}{k_0} \right)^2 \right]$$ 
with $k_{0} = c/ D_h$, 
 and $\omega_0 = c^2/D_h$, we see that there are 
two limiting cases :

\begin{itemize}
\item for $k \gg  k_{0}$,  
 $ S^{-1}(k,\omega=0) \sim k^{4}$; using again 
 $S^{-1}(k=0,\omega) \sim \omega^{2}$,
we obtain   $\alpha=1/2$ and $\beta=1/4$, $z=2$ via
Eqs.\ref{eq-kw1}.  
\item for $k \ll k_{0}$, $ S^{-1}(k,\omega=0) \sim
k^{2}$;using the fact that
the limit $S^{-1}(k=0,\omega)$ is always $\omega^{2}$, this
is consistent with the set of exponents
 $\alpha=0 $, $\beta=0$ and $z=1$ via
Eqs.\ref{eq-kw1}. 
\end{itemize}

The first of these contains no surprises, being the normal EW fixed point,
\cite{kn:ew}
while the second represents a new,
 `smoothing' fixed point.  

We now explain this smoothing fixed point with a 
  simple physical
picture. 
The competition between the two terms in Eq.\ref{eq-few}
determines the nature of the fixed point observed:
when the diffusive term dominates the flow term,
the canonical EW fixed point is obtained, in the limit
of large wavevectors $k$. On the contrary, when the
flow term predominates, the effect of diffusion
is suppressed by that of a travelling wave
whose net result is to penalise large slopes; this
leads to the smoothing fixed point obtained
in the case of small wavevectors $k$.

We add in passing that this crossover would not have
been as transparently obvious had we been using
the single Fourier transforms
$S(k,t=0)$ and 
$S(x=0,\omega)$
for  numerical purposes. We illustrate this by writing
explicitly the  expressions for the relevant
quantities:
\begin{eqnarray}
S(k,t=0)      &\sim  &{1\over k^{2} }\\
S(x=0,\omega) &\sim   &{1\over \omega^{2}} \hskip 0.5cm \mbox {for $\omega$ small}  \label{eq-w2}\\
S(x=0,\omega) &\sim   &{1\over \omega^{1.5}} \hskip 0.5cm \mbox {for $\omega$
large} \label{eq-w1.5}
\end{eqnarray}

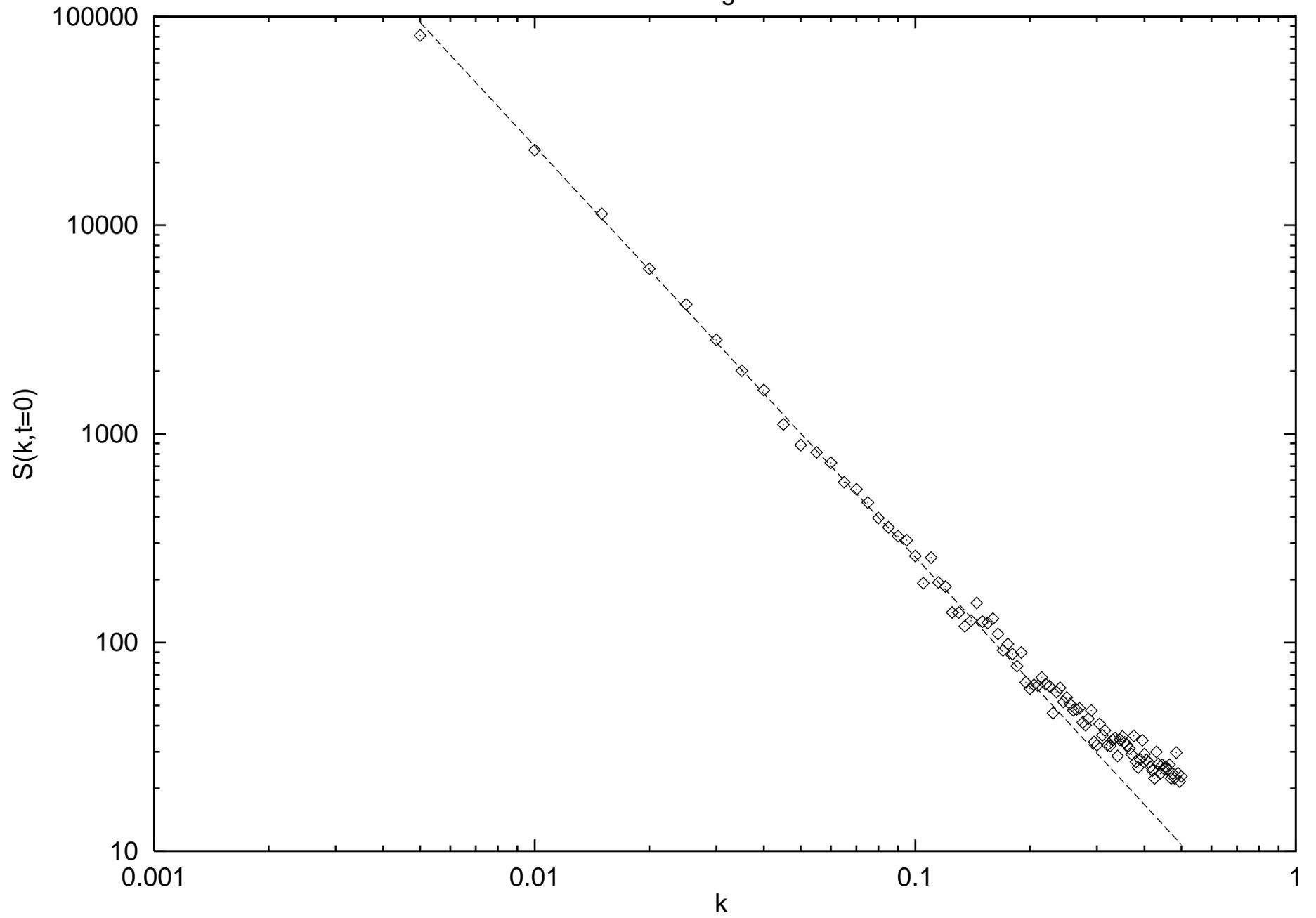
\begin{figure}
\caption{
The single Fourier transform $S(k,t=0) $
for the $h$-$h$ correlation function, along
with a power-law fit to  $k^{-1.97 \pm 0.02}$. 
}
\label{fig-4}
\end{figure}

The examination of $S(k,t=0)$  (Fig.\ref{fig-4}) on its own yields
no indication of the crossover to the smoothing
fixed point, while only a detailed examination
of the
 $S(x=0,\omega)$ graph (Fig.\ref{fig-5}a) reveals this, indicating a
crossover from $\omega^{-1.5}$ at high frequencies to $\omega^{-2}$
at low frequencies.
Clearly, though 
 this is more time-consuming to obtain, the double Fourier
transform, ie the full structure factor,
 provides a much more unambiguous picture of the
crossover in this problem.

Additionally, the investigation of even the single Fourier transform
to detect the crossover is not without its complexities.
The detection of the smoothing fixed point, characterised by  $z=1$ ,
needs a careful investigation of the 
 low frequency part of $S(x=0,\omega)$.
This is elaborated on in Appendix~\ref{app-osc},
where we discuss the anomalous oscillatory behaviour obtained
to do with the competition between grid sizes
and some
 characteristic lengths in this problem 
 (Appendix~\ref{app-scl}).

In view of the above, it is preferable to use the double Fourier
transform to obtain an unambiguous picture of 
the structure factor although this strategy might on first appearance
seem to be a computational overkill.  The overwhelming advantage is
that, by scanning the structure factor as a function of frequency
$\omega$ for a fixed $k$, one immediately sets two frequency scales $ck$ and $D_h k^{2}$, thus
making it possible to pick up the relevance of these scales in $S(k,\omega)$.

Before turning to an interesting physical application of the two fixed points in 
our linear equation, we mention in passing that our discussion is equally applicable
to the Kardar-Parisi-Zhang (KPZ) equation \cite{kn:kpz} with the addition of a flow term. Here too,
the use of the double Fourier transform  reveals the presence of the `smoothing' 
fixed point due to the flow term \cite{4auth}.

\subsection{Coupled equations: a model of smoothing}
 
Returning to the coupled equations
(\ref{eq-ew}), we realise from the above that the interface $h$
is smoothed because of the action of the  flow term which 
penalises the sustenance of finite gradients $\na h$.
As we have seen from the above analysis, this is indeed the case for the equation in $h$,
which is effectively decoupled from the equation in $\rho$. However,
the equation in $\rho$ is manifestly coupled to $h$
so we would need to ensure that no instabilities
are generated in this,
in order for the coupled equations to qualify as a valid
model of sandpile dynamics.

In this spirit, we look first at the value of $\rho$ averaged over the pile,
as a function of time (Fig.\ref{fig-6}a). We observe that the incursions of $\langle\rho\rangle$ 
into negative values are limited to relatively small values, suggesting
that the addition of a constant background of $\rho$ exceeding this negative value would 
render the coupled system meaningful, at least to a first approximation.
In order to ensure that this average over the pile does 
not involve wild fluctuations over the body
of the pile, we examine the fluctuations in $\rho$,
viz.$ \sqrt{\langle\rho^{2}\rangle - \langle\rho\rangle^{2}}$ (Fig.\ref{fig-6}b). The trends in that
figure indicate that this quantity appears to saturate,
at least upto computationally accessible times.
Finally we look at the {\it minimum} value of
$\rho $ at any point in the pile over a large range
of times (Fig.\ref{fig-6}c);  this appears to be
bounded by a modest (negative) value of `bare' $\rho$.
Our conclusions are thus that the 
fluctuations in $\rho$ saturate at computationally accessible
 times  and that the
negativity of the fluctuations in $\rho $ can always
be handled by starting with a constant $\rho_0$, a
constant `background' of flowing grains, which is
more positive than the largest negative fluctuation.

Physically, then, the above implies that at least in the presence of a constant large density $\rho_0$ of
flowing grains, it is possible to induce the level of smoothing corresponding
to the fixed point $\alpha = \beta = 0.$ This model is
thus one of the simplest possible ways in which
one can obtain a representation of 
 the smoothing that is frequently observed
in experiments on real sandpiles after  avalanche 
propagation \cite{kn:sdnagel}.

\section{ CONCLUSIONS}

In conclusion we have presented a very simple model
of sandpile dynamics which manifests the
interfacial smoothing commonly
observed after avalanche propagation,
both in experiments \cite{kn:sdnagel}
and physically in sand dunes. This
consists simply of a set of coupled continuum
equations \cite{kn:ambook,kn:mln} which model the exchange
process between the stuck grains at the
interface $h$ and the flowing grains $\rho$
which flow across the interface. The
equation in $h$ is closely related
to the well-known Edwards-Wilkinson equation
\cite{kn:ew}, modifying this simply by the addition
of a flow term which penalises
excess gradients across the interface,
and thereby causes the smoothing.

We find, on analysing the equations,
that there are two fixed points
which characterise the interface equation:
one, the normal diffusive fixed point
corresponding to
 $\alpha=1/2$, $\beta=1/4$ and $z=2$,
and the other, the new smoothing
fixed point corresponding to
 $\alpha=0 $, $\beta=0$ and $z=1$.
We mention here that the first
of these is the one that would also
be observed in the frame of the avalanche
moving with the flow velocity $c$,
while the smoothing would only be
observed in the stationary frame
of an observer, who would observe
the smoother surface left in the wake of
the avalanche.

Finally, in work that is in progress,
we have observed quantitatively
this smoothing phenomenon
in a cellular automaton model of
sandpiles \cite{kn:garywidth}. We
hope that this will motivate
actual experiments to measure
more quantitatively
this so far qualitatively
observed and intuited
behaviour at sandpile surfaces.

\section*{Acknowledgements}

Arnab Majumdar acknowledges the  hospitality of SNBNCBS
during the course of this work, while Anita Mehta acknowledges valuable discussions 
with Prof. Sir Sam Edwards during its initiation.

\appendix

\section{Oscillations} \label{app-osc}

We explain here the origins of the observed 
 oscillations in the structure factor
 $S(x=0,\omega )$ . 
The single Fourier transform $S(x,\omega )$ is defined by 
\be
S(x=0,\omega) = {1\over 2\pi} \int_{-\iff}^{\iff} S(k,\omega)exp(ikx)dk
\ee
In the limit of small $\omega$ the integral can be written as:  
 
\be
S(x=0,\omega) = {1\over 2\pi} \int_{-\iff}^{\iff}{1\over D k^{2}} \delta(\omega-ck) dk 
\ee

This is the origin of the ballistic behaviour
of the flow term and is responsible in part
for some of the observed oscillations,
as we explain below.
It is clear from Eqs.\ref{eq-w2}  and \ref{eq-w1.5} 
that the crossover  
 from small
to large $\omega$ for $x =0$ should not involve
any imaginary quantities, and therefore strictly
speaking we should not see any oscillatory
behaviour in the structure factor in this limit.
However it is important
to realise that the full form of the structure
factor $S(x,\omega)$ for finite $x$
{\it does} contain imaginary portions,
in order to understand fully the origin of
the obtained oscillations.

The characteristic length and time scales in our
problem are given by  (Appendix~\ref{app-scl}) 
\\
$$ t_{0} = {D\over c^{2}} \mbox{\hskip 0.1cm and \hskip .1cm }  x_{0} = {D \over c}$$
Whenever grid sizes in time or space
are comparable to these characteristic
lengths, the profile fluctuates
across these intervals, which
is then aggravated by the shock
fronts associated with the flow term.
This results in:
\begin{itemize}
\item oscillatory behaviour
arising from the {\it non-zero} intervals in $x$ 
associated with the sampling of the profile to generate the Fourier
transform, $S(x=0,\omega)$,  which introduce a flavour
of $S(x,\omega)$ for {\it finite} $x$.
\item oscillations which become
increasingly violent as $c$ increases
because of the increased fluctuations
associated with the ballistic
flow term over the grids.
\end{itemize}

This line of reasoning  is borne out by the following table:
\begin{tabbing}
2222222\=2222222222222222222222\=22222222222222222222222222\=\kill
\>{\bf Parameters} \> {\bf Grid Sizes \&}\> {\bf Observation  }      \\
\> \> {\bf Characteristic  Lengths}\>     \\
\> \> \> \\
\> c=1  $D_h$ = 1  \> $\Delta x=0.5, x_{0}=1.0 $ \>  No Oscillation . \\
\>                 \> $\Delta t=0.001, t_{0}=1.0 $ \>Fig.\ref{fig-5}a             \\
\> \> \> \\
\> c=5  $D_h$ = 1  \> $\Delta x=0.1, x_{0}=0.2 $ \>  Oscillation(1 peak). \\
\>                 \> $\Delta t=0.001, t_{0}=0.04 $ \> Fig.\ref{fig-5}b      \\
\> \> \> \\
\> c=10  $D_h$ =1  \> $\Delta x=0.1 ,x_{0}=0.1 $ \>  Oscillation( 2 peaks). \\
\>                 \> $\Delta t=0.001, t_{0}=0.01 $ \>Fig.\ref{fig-5}c              \\
\end{tabbing}

Thus in order to avoid these oscillations,
one should choose grid sizes
$\Delta x $ and $\Delta t $ in such a way that they
are always less than
the characteristic scales in the problem, i.e.,

  $$\Delta x \ll x_{0} \quad \mbox{ and } \quad \Delta t \ll t_{0} .$$ 
 
 \section{Length and time scales of interest}\label{app-scl}

  The Edwards-Wilkinson equation with a flow term ($c\nabla h$) 
  is given by
  \begin{equation}
   \frac{\partial h}{\partial t} = D_h \nabla^2 h + c \nabla h + \eta(x,t)
  \end{equation}
  with
  \[ \langle \eta(x,t) \eta(x^{\prime},t^{\prime})\rangle = \Delta^2
   \delta(x - x^{\prime})\delta(t - t^{\prime}) \]
  We perform a scale change in order to make this equation dimensionless
  using 
  \[ t = t_0 T \, , \quad x = x_0 X \, , \quad h = h_0 H  \]
  and obtain
  \begin{equation} 
  \frac{\partial H}{\partial T} = D_h^{\prime} \frac{\partial^2 H}{\partial X^2}
   + c^{\prime} \frac{\partial H}{\partial X}+ \delta_0 \xi
  \end{equation}
  where,
  \[
    D_h^{\prime} = D_h \frac{t_0}{x_0^2} \, , \quad
    c^{\prime} = c \frac{t_0}{x_0} \, , \quad
    \delta_0 = \frac{\Delta}{h_0} \sqrt{\frac{t_0}{x_0}} 
  \]
  and
  \[ \langle \xi(X,T) \xi(X^{\prime},T^{\prime})\rangle =
     \delta(X - X^{\prime})\delta(T - T^{\prime}) \]
  Next we make a choice of the scales ($x_0$, $t_0$ and $h_0$) such that the
  coefficients $D_h^{\prime}$, $c^{\prime}$ and $\delta_0$ are unity. This
  gives the following scales 
  \begin{equation}
   t_0 = \frac{D_h}{c^2} \, , \quad
   x_0 = \frac{D_h}{c} \, , \quad
   h_0 = \frac{\Delta}{\sqrt{c}}
  \end{equation}

  An estimate of the exponent $z$ can be made by considering the diffusion and
  flow term separately. 
  Substituting $L$ for $x_0$ along with the condition $D_h^{\pr} =
  c^\pr = 1$ we realise that 
  on length scales comparable to the system size, $L$, 
  the associated time scale for diffusion is $L^2/D_h$ and that for flow is
  given by $L/c$. The physical meaning of these time scales is the time taken 
  for the diffusive and flow-like correlations to span the system. This is 
  the time ($t_{\times} \sim L^z$) for crossover to the saturated regime. Clearly 
  the above time scales indicate
  $z = 2$ for diffusion-dominated behaviour and $z=1$ for flow-dominated 
  behaviour respectively.

\begin{figure}
\caption{
(a)The single Fourier transform $S(x=0,\omega)$
for the $h$-$h$ correlation function, along
with a power-law fit to ${\omega^{-1.97 \pm 0.05}}$
 for small $\omega$ and ${\omega^{-1.48 \pm 0.02}}$
for large $\omega$.\\ 
(b) and (c) show the 
oscillatory behaviour (see Appendix~\ref{app-osc})  
of 
$ S(x=0,\omega)$ near the crossover region.
The oscillations increase with increasing 
 $c$ ,  as can be seen from a comparison of Fig. (b)
($ c=10$) and Fig. (c) ($c=5$).
$D_h = \Delta^2 = 1.0$ throughout.
}
\label{fig-5}

\end{figure}

\begin{figure}
\caption{
Curves relating to the behaviour of the equation in $\rho$ (Eq.\ref{eq-ew})
over $10^6$ timesteps.\\ 
(a) shows the behaviour of the mean ${\langle \rho \rangle}$ as
a function of time, where the average
over the sandpile was computed over a sample
configuration.\\
(b) shows the variance $\rho_{rms} $ of $\rho$ as a function of time.
This was computed for 100 configurations over
the sandpile surface.\\  
(c) shows the bounds
of the fluctuations in $\rho$, $\rho_{min}$ and 
$\rho_{max}$, as a function of time. 
}
\label{fig-6}
\end{figure}

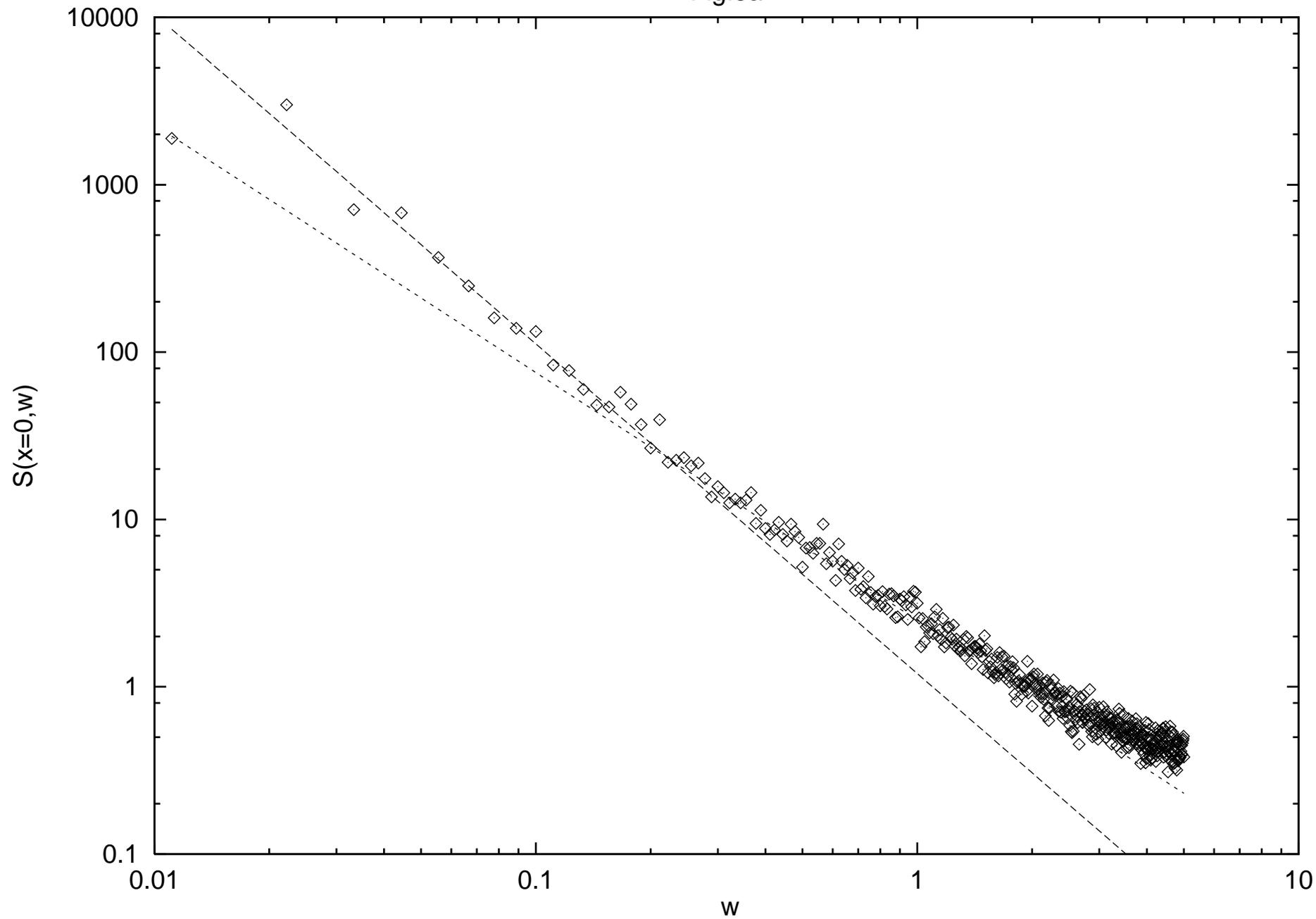
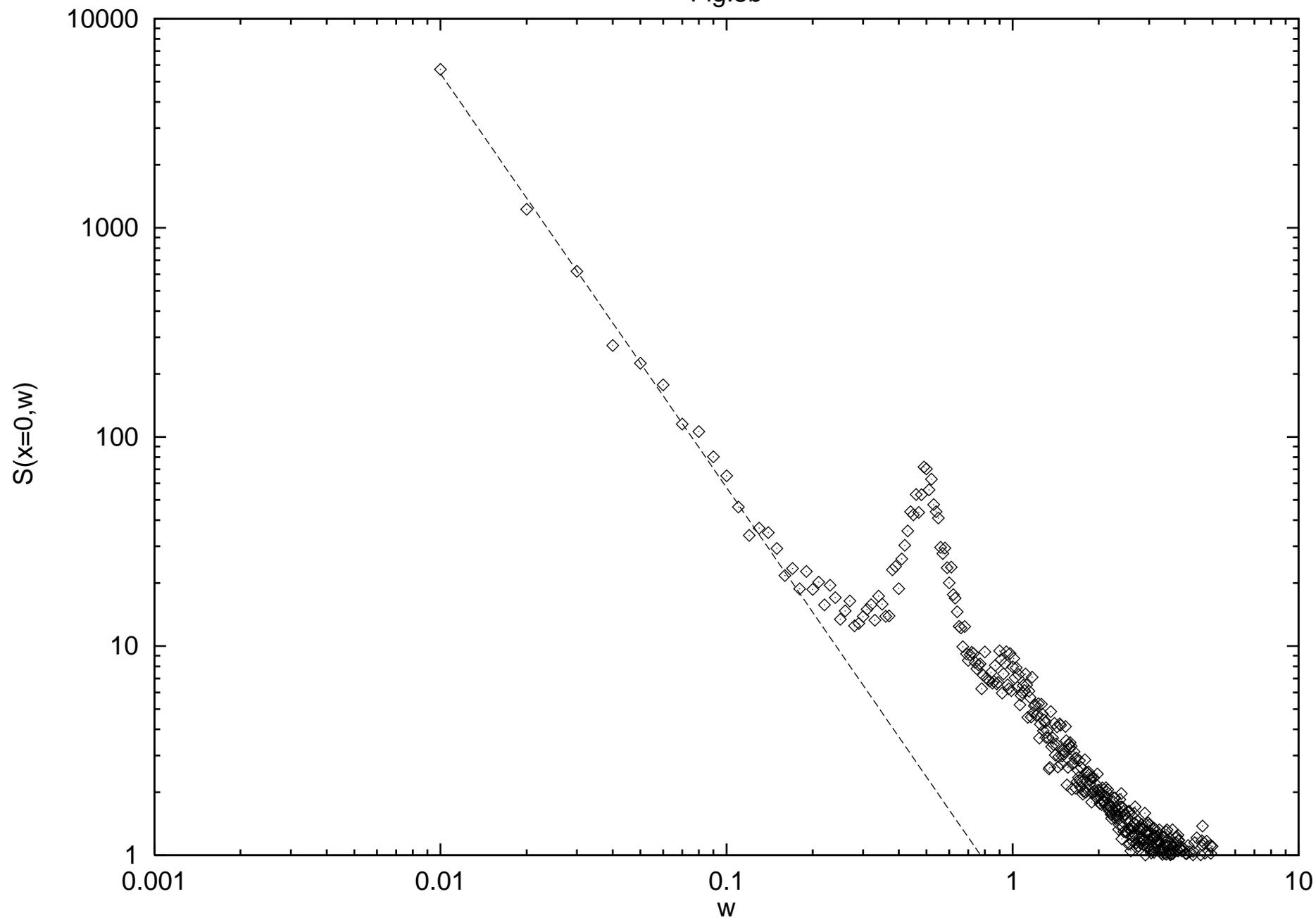
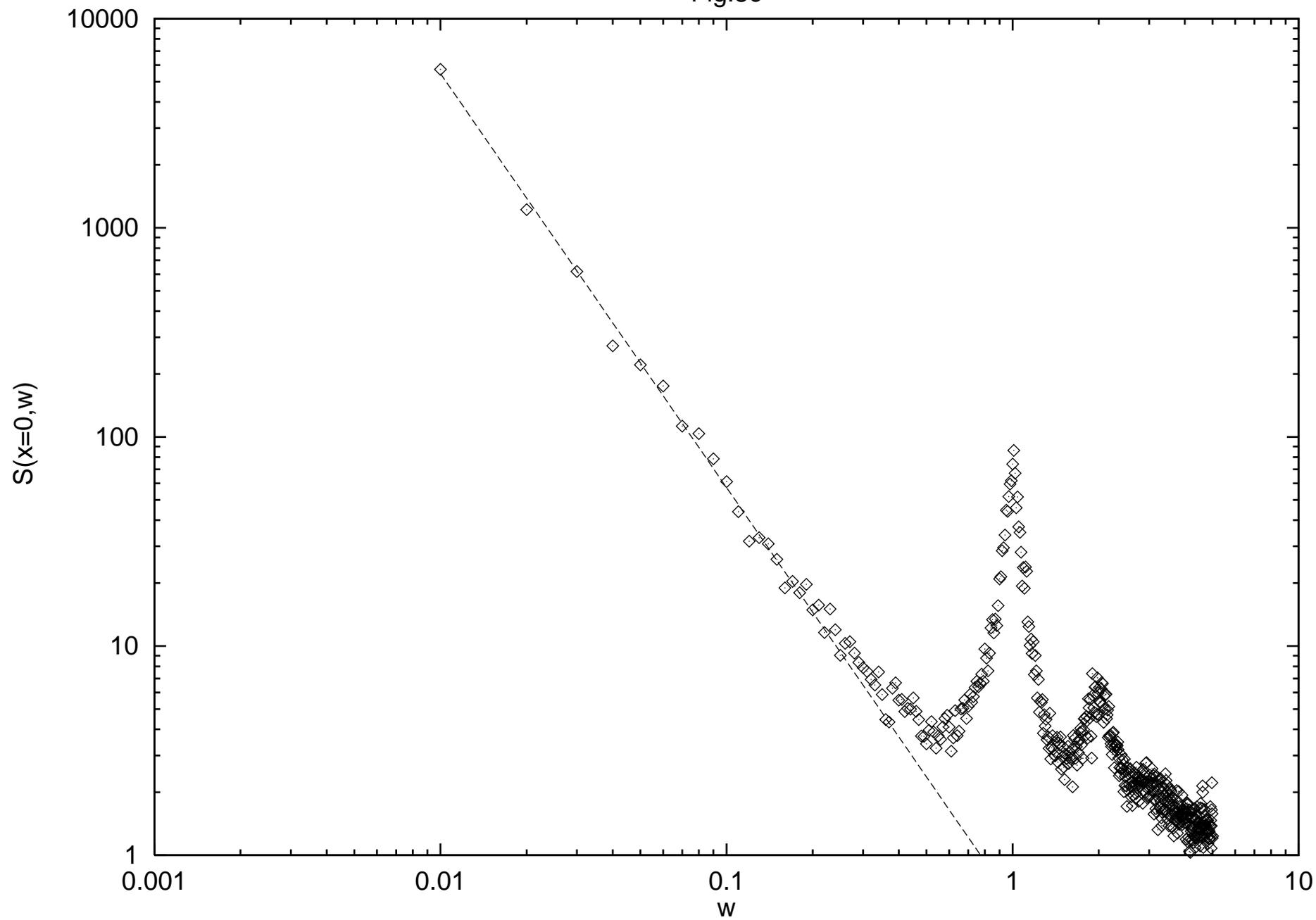
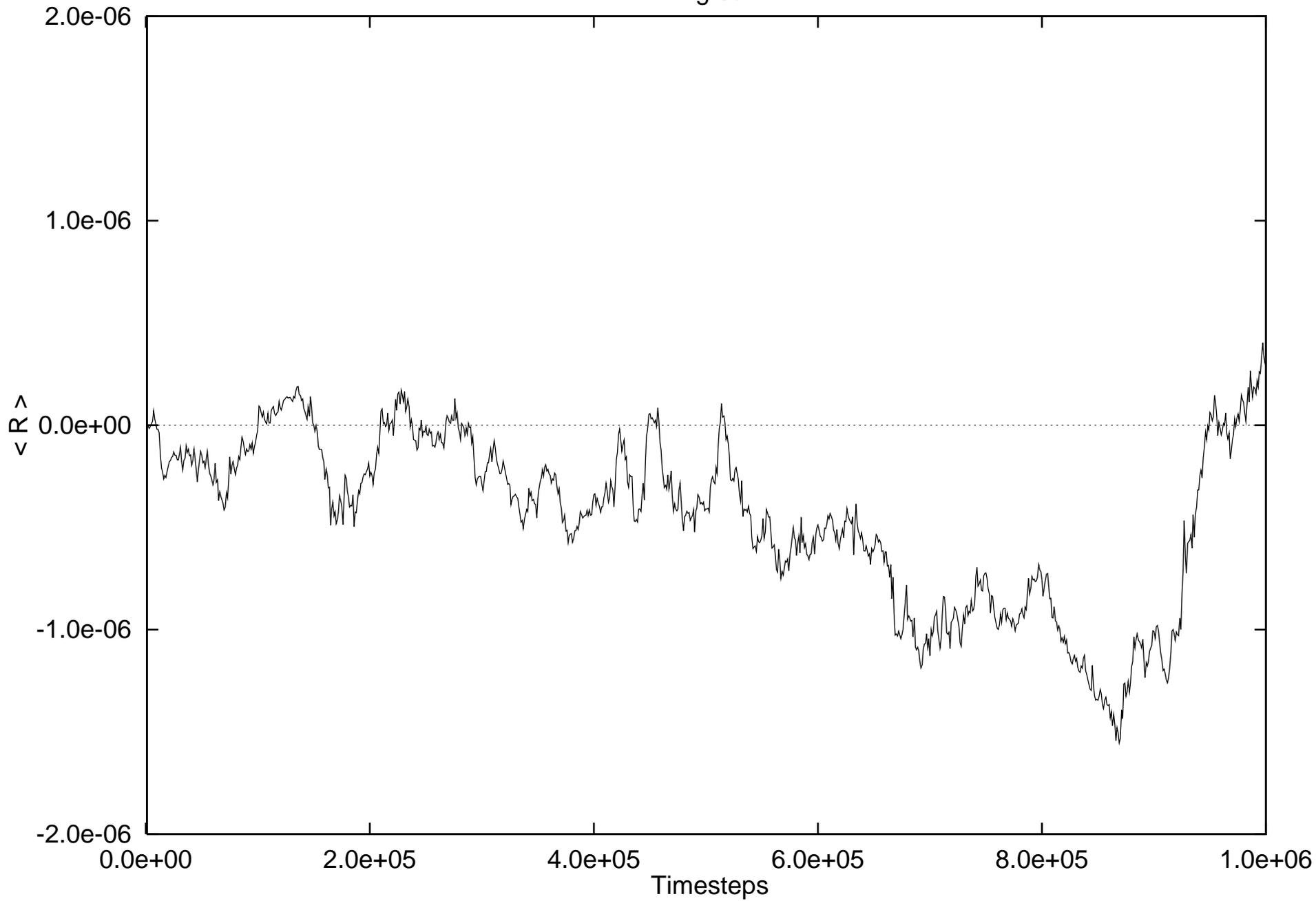
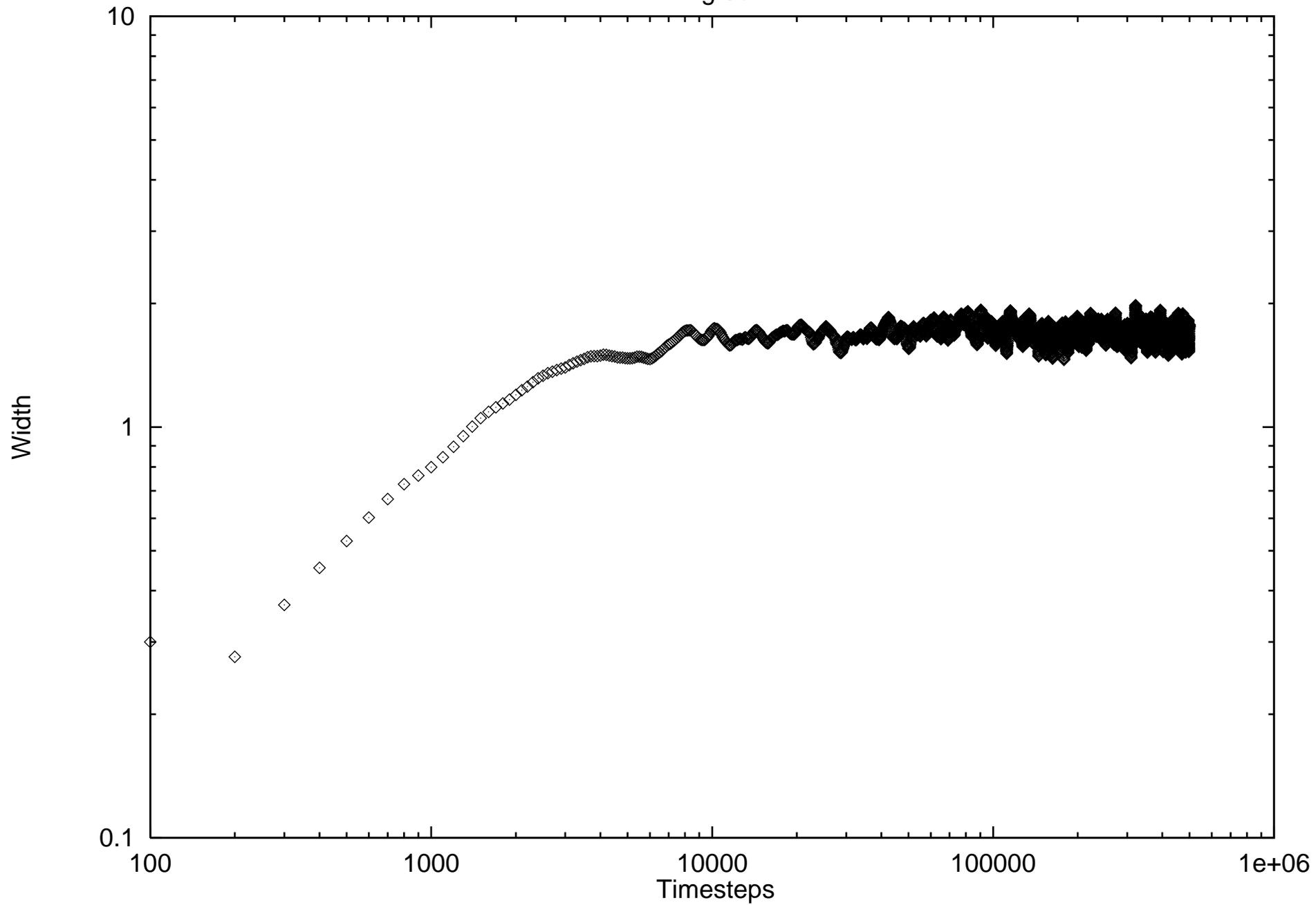
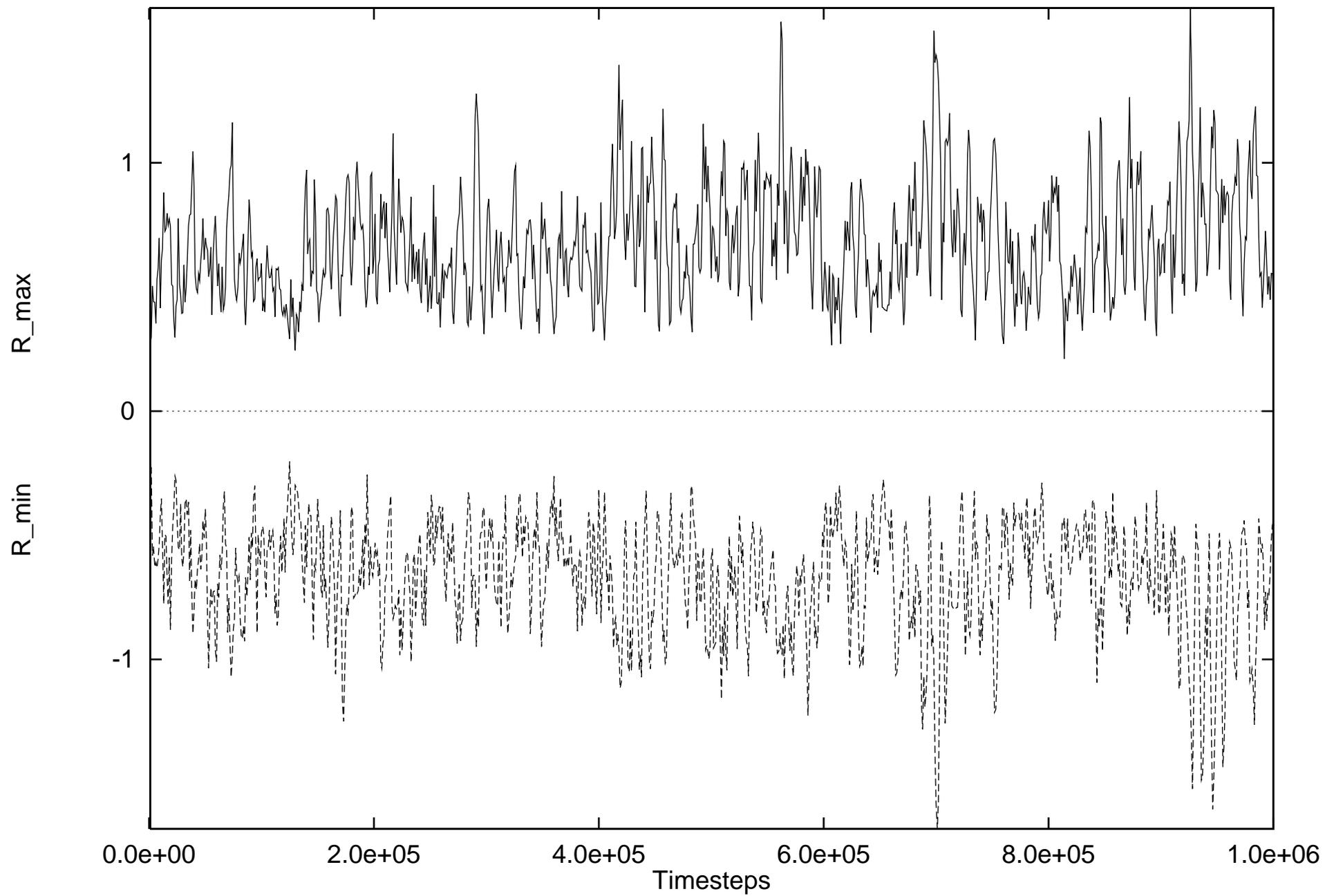
\end{document}